\documentclass[twocolumn,showpacs,preprintnumbers,amsmath,amssymb,a4paper,superscriptaddress]{revtex4}

\usepackage{graphicx}
\usepackage{dcolumn}
\usepackage{bm}

\begin{document}

\title{Tunable transmission via quantum state evolution in oval quantum dots}

\author{D. Buchholz}
\affiliation{Theoretische Chemie, University of Heidelberg, INF 229, 69120 Heidelberg, Germany}
\author{P. Drouvelis}
\affiliation{Tyndall National Institute, Lee Maltings, Prospect Row, Cork, Ireland}
\author{P. Schmelcher}
\affiliation{Theoretische Chemie, University of Heidelberg, INF 229, 69120 Heidelberg, Germany}
\affiliation{Physikalisches Institut, University of Heidelberg, Philosophenweg 12, 69120 Heidelberg, Germany}

\date{\today}

\begin{abstract}
We explore the quantum transmission through open oval shaped quantum dots. 
The transmission spectra show periodic resonances and, depending on the geometry parameter, 
a strong suppression of the transmission for low energies.  
Applying a weak perpendicular magnetic field changes this situation drastically and introduces a large conductance. 
We identify the underlying mechanisms being partially due to the specific shape of the oval that causes a systematic 
decoupling of a substantial number of states from the leads. 
Importantly a pairwise destructive interference of the transmitting states is encountered 
thereby leading to the complete conductance suppression. 
Coupling properties and interferences can be tuned via a weak magnetic field. 
These properties are robust with respect to the presence of disorder in the quantum dot.
\end{abstract}

\pacs{73.23.-b,73.23.Ad,75.47.Jn}

\maketitle

Magnetoconductance of two-dimensional mesoscopic structures in semiconductors is 
an intense field of current research both with respect to its theoretical understanding 
as well as possible applications  \cite{Akis1997,Datta1995,Nogaret2000,Marcus1992,Baranger1993,Chang1994,Sachrajda1998,Marlow2006,Timp1987,Olendski2005,Bird2003,Elhassan2004,Sadreev2004,Klitzing1980,Laughlin1981}. 
The quantum Hall effect \cite{Klitzing1980,Laughlin1981} and its various applications, 
are spectacular examples for high magnetic field strengths. 
In the regime of weak magnetic fields  
phenomena like 
weak localization \cite{Marcus1992,Baranger1993,Chang1994} 
and fractal conductance fluctuations \cite{Marcus1992,Jalabert1992,Sachrajda1998,Marlow2006} 
in quantum billiard systems and the Aharonov-Bohm effect in, e.g., quantum rings \cite{Timp1987} represent important features. 
For higher energies scarring effects are observed \cite{Akis1997}, 
which can be described semiclassically \cite{Wirtz2003,Jacquod2006} in many cases. 
Semiconductor nanostructures have shown to be a testing ground for fundamental physics models 
and allow to investigate the quantum to classical crossover. 
Apart from this they are the building elements of future quantum based electronics.
Here coherent control of electronic states is a necessity for the integration of quantum effects. 
Beyond generic effects due to disorder and chaos, 
the specific shape of the confining potential has proven to be of great importance.  
The magnetoconductance of curved quantum waveguides, for example, depends strongly on the bending \cite{Olendski2005}. 
Arrays of rectangular quantum dots show a metal to insulator transition under application of magnetic fields \cite{Elhassan2004}. 
In certain cases transmission through open systems can be determined by only few eigenstates of the closed system \cite{Bird2003}. 
Making use of these properties for designing conductance is highly desirable. \\
In this letter, we focus on the deep quantum regime of low energies and weak magnetic fields and 
explore the quantum transmission through an open, oval shaped billiard system 
in the ballistic regime. 
Our approach is based on the single particle picture where 
effects of electron-electron and electron-phonon scattering are neglected. 
Experimentally this may be assured by reducing the temperature and the system size 
to the regime where inelastic scattering has no significant impact \cite{Steinbach1996,Henny1999,Oberholzer2001}. 
For circular or rectangular billiards it is well known that the corresponding transmission spectra possess a strongly 
fluctuating character \cite{Sadreev2004}. 
In contrast, the obtained transmission spectra for the oval exhibit a highly regular behavior. 
Depending on the degree of deformation, 
the transmission can be completely suppressed for large parts of an entire channel  
or can become maximal. 
A weak perpendicular magnetic field drastically changes the 
transmission characteristics from small to large values and vice versa. 
The suppressed conductance is caused by the decoupling of many 
eigenstates from the leads 
accompanied by systematic degeneracies and destructive interference 
for a certain class of states.
The magnetic field changes the coupling strengths and the relative phases 
of these states resulting in a magnetic switching effect for very weak field strengths.
The conductance suppression as well as the switching effect are present even in the case of significant disorder.\\
The oval \cite{Berry1981} is parametrized as follows: 
\begin{eqnarray}
x(\phi) & = & R [(1 + \frac{\delta}{2}) \sin(\phi) + \frac{\delta}{6} \sin(3\phi)] \nonumber \\ 
y(\phi) & = & R [(-1 + \frac{\delta}{2}) \cos(\phi) - \frac{\delta}{6} \cos(3\phi)]
\end{eqnarray}
with $\phi \in [0,2\pi[$. The parameter $\delta$ defines the deformation of the oval. 
For $\delta = 0$ the boundary describes a circle, and for $\delta > 0$ one obtains an elongated shape. 
Semi-infinite leads of width $W=0.3 R$ are attached to the right and left side of the oval. 
Whenever employing SI-units, we refer to an exemplary choice $R=220$ nm. 
A reduction of the length scale $R$ would lead to an increase of energy and 
magnetic field strength $\sim \frac{1}{R^2}$. 
The Hamiltonian reads $H = \frac{(\vec{p} - e\vec{A})^2}{2 m_{\mathrm{eff}}} + V(\vec{r})$, 
where the potential $V(\vec{r})$ is chosen to be zero 
inside the leads and the oval and infinite outside.  
The effective mass is chosen to be 
$m_{\mathrm{eff}} = 0.069 m_e$ where $m_e$ is the electron mass. 
$\vec{A}$ is the magnetic vector potential of a homogeneous perpendicular magnetic field. 
The electronic spin is not taken into account since 
for the considered regime of field strengths $ < 0.1$T the Zeeman spin splitting in GaAs 
($g_{\mathrm{eff}} = 0.44$) is less than $0.0025$meV. \\
The Hamiltonian is discretized on a tight-binding grid  
using the Peirls' substitution for the vector potential $\vec{A}$. 
The semi-infinite leads on the left and right side in the absence of a magnetic field
are provided by analytic expressions for the self-energies ${\bf \Sigma}_{l/r}$, as described e.g. in \cite{Datta1995, Ferry1997}. In the case of a nonzero magnetic field, the field strength is linearly decreased to zero within a sufficiently long region of the leads. We have validated the independence of our results on the length of this region.
The single particle Green's function is given by a matrix equation: 
 ${\bf G}(E) =[ E{\bf I} - ( {\bf H} + {\bf \Sigma}_r  + {\bf \Sigma}_l ) ]^{-1}$  
and the transmission is obtained \cite{Datta1995} via 
$T(E) = Tr \left({\bf  \Gamma}_r {\bf G} {\bf \Gamma}_l {\bf G}^\dagger \right)$
with ${\bf \Gamma}_{l/r} = i ( {\bf \Sigma}_{l/r} - {\bf \Sigma}^\dagger_{l/r})$. 
We use a parallel implementation of the recursive Green's functions method for the calculation of the 
transmission \cite{Drouvelis2006}. 
Eigenstates and eigenenergies of the closed system are obtained via a block Lanczos algorithm. \\
\begin{figure}
 \includegraphics[width=\columnwidth]{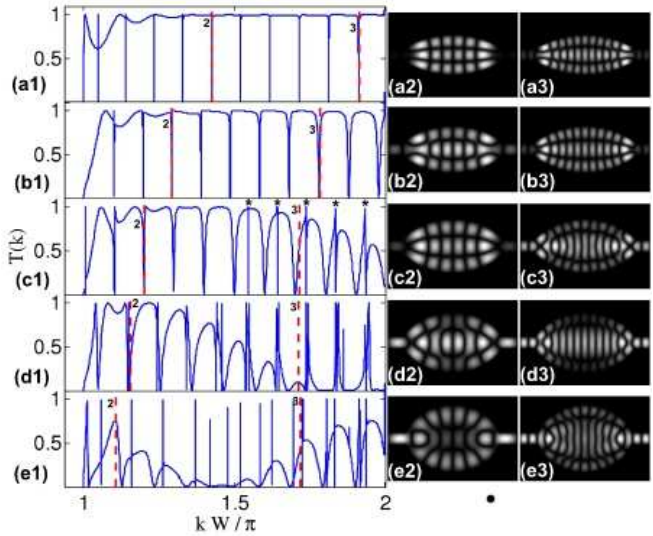}
 \caption{\label{abb_1}(Color online) (a1-e1): Zero-field transmission through the open oval for different values of the deformation parameter 
 (a) $\delta = 1.5$; (b) $\delta = 1.25$; (c) $\delta = 0.99$; (d) $\delta = 0.75$; (e) $\delta = 0.5$.
 (a2-e2): Density of the $(6,3)$ state (notation see text) in the closed oval including a part of the lead. 
 (a3-e3): Density of the $(11,3)$ state.  
 The $k$-positions of these states are indicated by the red-dashed lines, labeled 2 and 3 respectively, 
 in the according transmission spectra on the left. 
 }
\end{figure}
In Figs.~\ref{abb_1}.a1 to \ref{abb_1}.e1 we present the transmission 
as a function of the wavenumber $k = \frac{\sqrt{2m_{\mathrm{eff}} E }}{\hbar}$
for zero magnetic field and 
several values of the deformation parameter. 
The considered range of wavenumbers is the first channel $\frac{\pi}{W} < k < \frac{2\pi}{W}$, 
i.e. the propagating wave in the leads is in the transversal ground state. 
For the above specified system size this corresponds to $E \in [1.2, 5]$ meV. 
The transmission $T$ of the oval shows, for the presented range of values of $\delta$, 
less fluctuations in comparison with the calculations of Rotter {\it et al.}\cite{Rotter2003} for circular and stadium billiards (with leads oriented perpendicular to each other) and in comparison to our own results for rectangular and circular geometry.
For $\delta= 1.5$, $T$ is almost everywhere maximal, 
distracted only by one series of nearly periodic and extremely narrow resonance dips in the transmission.
Decreasing $\delta$ changes the transmission in the following way: 
The existing resonances respectively transmission dips are shifted and broadened and 
additionally further resonances occur. 
We remark that there are resonances possessing a width $<0.005$meV, which 
are not resolved in Fig.~\ref{abb_1}. 
However, these resonances do not contribute significantly to the conductance 
if thermal broadening is taken into account. 
Apart from individual resonances $T$ overall decreases with increasing value of $\delta$: 
For $\delta = 0.75$ and $\delta = 0.5$ the transmission is suppressed in the upper 
and in the central third of the first channel, respectively. 
In order to explain the small number of resonances and the suppression of the transmission, 
we will analyze how the eigenstates of the closed oval (including short lead stubs)
contribute to the transmission when the system is connected to the leads. 
\begin{figure}
 \includegraphics[width=\columnwidth]{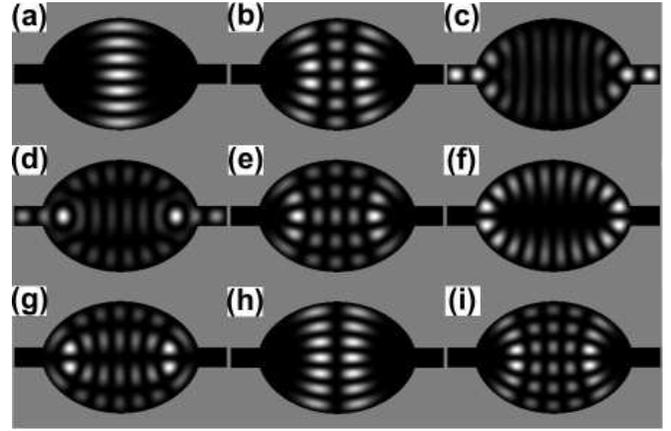}
 \caption{\label{abb_3} Densities of the 41st to 48th (a-h) and the 59th (i) eigenstate 
 in the closed oval with $\delta = 0.5$ for $B = 0$ in energetical ascending order. 
 The states are (a) $(1,7)$; (b) $(3,6)$; (c) $(11,1)$; (d) $(8,3)$; 
 (e) $(5,5)$; (f) $(10,2)$; (g) $(7,4)$; (h) $(2,7)$; (i) $(5,6)$.}
\end{figure}
Let us first 
introduce a nomenclature for the states similar to that for rectangular cavities: 
We simply count the number of nodal domains (a nodal domain is a connected region, where the wavefunction possesses the same sign) in the longitudinal $x$- and in the transversal $y$-direction, 
using the notation $(n,m)$ for a state with $n$ nodal domains in the $x$- and $m$ 
in the $y$-direction. 
The examples in Fig.~\ref{abb_3} show, in energetical ascending order, the 41st to 48th  (Figs. \ref{abb_3}.a to \ref{abb_3}.h) and the 59th eigenstate (\ref{abb_3}.i) of the oval with $\delta = 0.5$. 
Of course, we consider only the nodal domains that are located inside the oval, excluding those in the lead stubs \cite{Fussnote1}. \\
The number of eigenstates lying in the energy range of the first channel 
varies form 65 for $\delta = 1.5$ to 100 for $\delta = 0.5$
but far fewer eigenstates (in the order of 10 to 20) contribute significantly to the zero-field transmission.
To understand this, we provide a qualitative analysis of the coupling  
strengths of individual eigenstates to the leads. 
At first, all states with even $m$ (e.g. Figs.~\ref{abb_3}.f, \ref{abb_3}.g and \ref{abb_3}.i) 
do not contribute to the transmission as they possess negative transversal $y$-parity, 
implying zero overlap with the propagating states in the leads, such that the zero field transmission is determined 
by the interplay of $(n,m)$ states with odd $m$.
Later we will see that the states with even $m$ make one contribution to the magnetoconductance for non-zero field strength.\\
In the following, we distinguish between confined states (CS) and leaking states (LS). 
A CS does not fill the whole area of the oval with a substantial probability amplitude  
but is located near the center $x = 0$ and decays rapidly in the direction towards the lead stubs 
(e.g. Fig.~\ref{abb_3}.a,b,e,h,i). 
The LS possess a high probability amplitude in the lead stubs (e.g. Fig.~\ref{abb_3}.c and d)
and accordingly a short lifetime in the open system. 
They are therefore expected to make a substantial contribution to the transmission within a broader energy range. 
The CS $(n,m)$ (with odd $m$) cause resonances possessing a width smaller than the average level spacing 
thus showing the typical Fano profile \cite{Fano1961}. 
For the CS we observe that increasing $n$ increases and increasing $m$ 
decreases the states longitudinal $x$-extension: 
see for example the $(2,7)$ state (Fig.~\ref{abb_3}.h) compared to the $(1,7)$ state (Fig.~\ref{abb_3}.a)
and the $(5,5)$ state (Fig.~\ref{abb_3}.e) compared to the state $(5,6)$ (Fig.~\ref{abb_3}.i), respectively. 
The latter can be understood intuitively as a higher transversal excitation (larger $m$)
hinders the wavefunction to penetrate into regions with smaller $y$-extension. 
For fixed $m$, one obtains a series of states $(n,m)$ for which the $x$-extension 
and accordingly the coupling to the leads increases with increasing $n$. 
This manifests itself in the transmission: 
In the case $\delta = 1.5$ (Fig.~\ref{abb_1}.a1) there is one series of resonances 
caused by the $(n,3)$ states with $n = 2,\ldots,11$.  
For example in Figs.~\ref{abb_1}.a2 and \ref{abb_1}.a3 the states $(6,3)$ and $(11,3)$, respectively, are shown. 
Their $k$-positions are marked with red dashed lines in the transmission spectrum labeled by '2' and '3' respectively. 
Both states are of the CS type. 
Resonances due to $(n,m)$ states with $m = 5$ are not resolved here, 
since increasing $m$ decreases the $x$-extension severely 
and accordingly the resonances' widths are much smaller. 
Aside the resonances, the transmission is governed by the LS. 
The LS of the closed system can be viewed as a superposition of two counterpropagating waves 
in the open system coming from the left and right lead. 
These waves obey a certain phase relation to fulfill the boundary conditions at the outermost walls of the lead stubs 
which can be extracted from the eigenstates. 
For $\delta = 1.5$, only the $(n,1)$ states are of the LS type. 
They cause a large transmission that is only modified by the sharp resonances due to the CS.\\
For $\delta =1.25$ (Fig.~\ref{abb_1}.b) the resonances are caused by $(n,3)$ states with 
$n = 4,\ldots,13$.
The increased width of the oval causes firstly a shift of the resonances to lower energies (see labels '2','3' in Fig.~\ref{abb_1}.b1) and secondly a larger resonance width as the $x$-extension of the $(n,3)$ states increases
(see Figs.~\ref{abb_1}.b2 and \ref{abb_1}.b3). 
Additionally, an increasing width of the resonances in the $(n,3)$ series with increasing $n$ can be observed, as the coupling strength increases with $n$. 
The non-resonant transmission is still determined by the $(n,1)$ states alone and accordingly high. 
For $\delta = 0.99$ the $(n,3)$ states are of the CS type for low energies, 
e.g. the $(6,3)$ state (Fig.~\ref{abb_1}.c2), 
and are of the LS type for higher energies, e.g. state $(11,3)$ (Fig.~\ref{abb_1}.c3)
is strongly coupled to the leads. 
Therefore the according resonances - visible as dips in the transmission - possess  
a very small width for lower energies whereas at higher energies the resonance widths are larger.
Here additional resonances due to the $(n,5)$ states with $n = 5,\ldots,9$ (marked by asterisks) are 
broad enough to be resolved. 
For $\delta = 0.75$ Fano resonances due to $(n,5)$, $(n,7)$ and $(n,9)$ states of the CS type are visible. 
All $(n,3)$ states are now of the LS type (see Figs.~\ref{abb_1}.d2, \ref{abb_1}.d3) and the interference 
of $(n,1)$ and $(n,3)$ states changes the non-resonant transmission severely: 
it is suppressed in the upper third of the first channel. 
For $\delta = 0.5$ we have again broad oscillations due to the $(n,3)$ states 
which are of the LS type (see Figs. \ref{abb_1}.e2 and \ref{abb_1}.e3)
and sharp resonances due to the $(n,m)$ states with $m \ge 5$. 
\emph{Within the central third of the first channel the transmission is strongly suppressed.} 
Besides the decoupling of many states from the leads, 
the underlying mechanism is as follows: 
The two types of LS make up two scattering channels 
that show constructive or destructive interference. 
\emph{Within the broad window of low transmission states of the $(n,1)$ and the $(n,3)$ series are pairwise quasi-degenerate
with the paired states possessing different $x$-parity, 
leading to destructive interference in the outgoing lead.} 
For example, 
the states $(11,1)$ (Fig.~\ref{abb_3}.c) and $(8,3)$(Fig.~\ref{abb_3}.d)
are located at $\frac{k W}{\pi} = 1.3442$ and $1.3452$, respectively. 
The phase relation between the leads is zero for the $(11,1)$ and $\pi$ for the $(8,3)$ state. 
This mechanism of decoupling of eigenstates and destructive interference leading to a suppression of conductance is 
not a peculiarity of the oval, but is robust with respect to moderate changes of the geometry. 
For example, we obtain almost the same non-resonant transmission 
for an ellipse possessing the same length and width as the oval. 
Finally we remark that further decreasing $\delta$ and approaching the limit of circular geometry 
causes more and more states to become LS, 
such that the transmission is determined by the interference of many states 
leading to large fluctuations with varying energy.\\
\begin{figure}
 \includegraphics[width=\columnwidth]{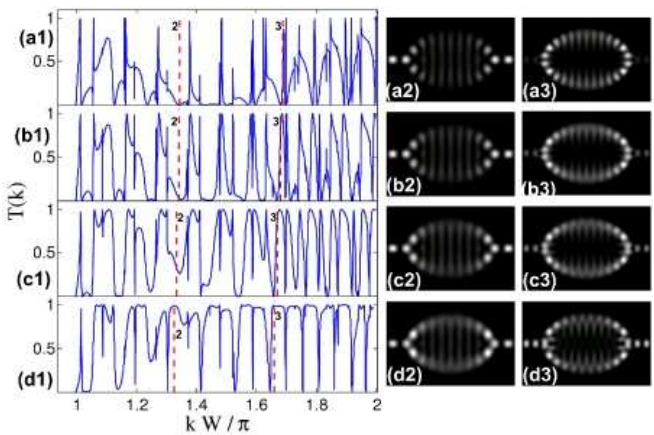}
 \caption{\label{abb_2}(Color online) (a1-d1) Transmission through the open oval with $\delta = 0.5$ for increasing 
 magnetic field strength. (a2-d2) Densities of the $(11,1)$ and (a3-d3) of the $(13,2)$ state in the closed oval. 
 The $k$-positions of these states are indicated by the red-dashed lines, labeled 2 and 3 respectively, 
 in the according transmission spectra on the left. 
 The magnetic field strength is (a) $B = 1.65$mT; (b) $B = 4.11$mT; (c) $B = 6.58$mT; and (d) $B = 9.46$mT.}
\end{figure}
Applying a weak perpendicular magnetic field changes the transmission drastically. 
In Figs.~\ref{abb_2}.a1 to \ref{abb_2}.d1 the transmission is shown 
for $\delta = 0.5$ and for several values of the magnetic field strength $B$.
Already for the smallest field strength many additional resonances occur. 
With increasing field strength some of these resonances broaden quickly. 
Accordingly, the average transmission rises, reaching a maximum of $T_{\mathrm{av}} > 0.8$ for $B=9.46$mT. 
The respective magnetic flux in units of $\Phi_0 = \frac{h}{e}$ through the oval is $\frac{\Phi}{\Phi_0} = 0.34$. Estimating the influence of thermal broadening by convolution of the transmission with the derivative of the Fermi function shows that suppression and enhancement of the conductance is stable at least for temperatures up to several Kelvin. 
For magnetic field strengths above $10$mT $T_{\mathrm{av}}$ decreases again and reaches a local minimum of $T_{\mathrm{av}} \approx 0.4$ at $B \approx 15$mT. For $B \approx 25$mT a local maximum of $T_{\mathrm{av}} \approx 0.7$ is reached. 
Height and position of these two local extrema vary moderately with the Fermi energy.
We encounter a second broad minimum of $T_{\mathrm{av}}$ for $B \approx 39$mT and further less pronounced minima and maxima for higher field strengths. 
Instead of a broad region of suppressed transmission there exists a narrow minimum whose position shifts to higher energies with increasing magnetic field strength.\\
The weak field necessary to cause a strong increase of the transmission 
does not substantially change the probability densities of the eigenstates of the corresponding closed system. 
This increase is caused by two mechanisms. 
(i) Although the approximate degeneracy of the $(n,1)$ and $(n,3)$ states in zero-field is only slightly lifted, 
these states acquire different Aharonov-Bohm phases, which prevents the destructive interference.
Figs.~\ref{abb_2}.a2 to \ref{abb_2}.d2 show the density of the $(11,1)$ state with  
its $k$-position being marked in the transmission curve by the label '2'. 
This state is quasi-degenerate with the $(8,3)$ state (Fig.~\ref{abb_3}.c). 
For $B=0$ transmission is nearly zero at this $k$-value. 
Increasing the magnetic field raises the transmission (see $T(k)$ in Fig.~\ref{abb_3}.a1-d1 at position '2'), 
while the phase difference between the two states at the right lead decreases.
(ii) The $(n,2)$ states that do not contribute to the zero-field transmission 
increasingly couple to the leads with increasing field strength. 
They lead to rapidly broadening resonances, resulting in an enhanced overall transmission. 
For example Figs.~\ref{abb_2}.a3 to \ref{abb_2}.d3 show the density of the $(13,2)$ state 
for different field strengths. 
With increasing field strength this state transforms to a LS, enhancing the conductance.\\
The symmetry related decoupling from the leads of the $(n,m)$ states with even $m$ does not suffice to explain the obtained magnetoconductance. 
With regard to symmetry, many states are allowed to contribute to the transmission. The circular and the rectangular billiard including their leads possess the same reflection symmetries and show neither any suppression of the transmission nor a remarkable change in the averaged $T$ for small values of the magnetic field strength.
Only if most of the $(n,m)$ states with odd $m$ are also decoupled from the leads according to the above described scenario, transmission can be blocked within a broad energy range and increased by a weak magnetic field.
The existence of further conductance minima and maxima with increasing magnetic field strength is the result of only few interfering channels acquiring different Aharonov-Bohm like phases.\\
We have studied the robustness of the above reported effects with regard to disorder and impurity scattering. 
Alternatively a random on site potential with amplitude $3$meV and an impurity potential with a mean free path of approximately $15 \mu$m have been used. 
The observed changes of the spectra due to disorder 
are the increase of the fluctuations of $T$, 
the shift of the exact positions of the resonances,
the occurrence of additional sharp resonances 
as well as a shift of the threshold for the onset of transmission to slightly higher energies.
However, we emphasize that in both cases of disorder 
all these changes are of minor extent and the overall behavior of the transmission, 
particularly the broad suppression of the conductance and the magnetic switching effect, 
remain largely unaffected, i.e. they are robust to disorder effects.\\
To conclude, we have shown that the oval represents a suitable set-up for conductance control via external magnetic fields. 
Its elongated, smoothly widening shape causes for zero field a systematic decoupling of most eigenstates from the leads.
Importantly, destructive interference of quasi-degenerate states results in an energetically broad suppression of the 
zero-field transmission which can be designed by the specific choice of the geometry parameters. 
Our studies show that this effect is not a peculiarity of the oval, 
but a systematic robust feature of many similar geometries that possess the same width to length ratio and a boundary with similar radius of curvature in the central region. 
A magnetic field introduces phases to the eigenstates that changes
their couplings to the leads thereby lifting the destructive interference effects. 
This allows to switch the transmission from small to large values and vice versa. 
The weak fields necessary to introduce major changes with respect to the conductance make our setup a promising 
candidate for a magnetically controlled mesoscopic transport element. 
We thank Peter Bastian, Mathias Brack and Roland Ketzmerick for very helpful discussions. 
D.B. acknowledges financial support of the Deutsche Forschungsgemeinschaft (DFG) within the IRTG 710. 
P.D. is thankful to the Irish Research Council for Science Engineering and Technology for funding.

\bibliography{./literatur}

\begin{thebibliography}{27}
\expandafter\ifx\csname natexlab\endcsname\relax\def\natexlab#1{#1}\fi
\expandafter\ifx\csname bibnamefont\endcsname\relax
  \def\bibnamefont#1{#1}\fi
\expandafter\ifx\csname bibfnamefont\endcsname\relax
  \def\bibfnamefont#1{#1}\fi
\expandafter\ifx\csname citenamefont\endcsname\relax
  \def\citenamefont#1{#1}\fi
\expandafter\ifx\csname url\endcsname\relax
  \def\url#1{\texttt{#1}}\fi
\expandafter\ifx\csname urlprefix\endcsname\relax\def\urlprefix{URL }\fi
\providecommand{\bibinfo}[2]{#2}
\providecommand{\eprint}[2][]{\url{#2}}

\bibitem[{\citenamefont{Akis et~al.}(1997)\citenamefont{Akis, Ferry, and
  Bird}}]{Akis1997}
\bibinfo{author}{\bibfnamefont{R.}~\bibnamefont{Akis}},
  \bibinfo{author}{\bibfnamefont{D.~K.} \bibnamefont{Ferry}}, \bibnamefont{and}
  \bibinfo{author}{\bibfnamefont{J.~P.} \bibnamefont{Bird}},
  \bibinfo{journal}{Phys. Rev. Lett.} \textbf{\bibinfo{volume}{79}},
  \bibinfo{pages}{123} (\bibinfo{year}{1997}).

\bibitem[{\citenamefont{Datta}(1995)}]{Datta1995}
\bibinfo{author}{\bibfnamefont{S.}~\bibnamefont{Datta}},
  \emph{\bibinfo{title}{Electronic Transport in Mesoscopic Systems}}
  (\bibinfo{publisher}{Cambridge University Press}, \bibinfo{year}{1995}).

\bibitem[{\citenamefont{Nogaret et~al.}(2000)\citenamefont{Nogaret, Bending,
  and Henini}}]{Nogaret2000}
\bibinfo{author}{\bibfnamefont{A.}~\bibnamefont{Nogaret}},
  \bibinfo{author}{\bibfnamefont{S.~J.} \bibnamefont{Bending}},
  \bibnamefont{and} \bibinfo{author}{\bibfnamefont{M.}~\bibnamefont{Henini}},
  \bibinfo{journal}{Phys. Rev. Lett.} \textbf{\bibinfo{volume}{84}},
  \bibinfo{pages}{2231} (\bibinfo{year}{2000}).

\bibitem[{\citenamefont{Marcus et~al.}(1992)\citenamefont{Marcus, Rimberg,
  Westervelt, Hopkins, and Gossard}}]{Marcus1992}
\bibinfo{author}{\bibfnamefont{C.~M.} \bibnamefont{Marcus}},
  \bibinfo{author}{\bibfnamefont{A.}~\bibnamefont{Rimberg}},
  \bibinfo{author}{\bibfnamefont{R.~M.} \bibnamefont{Westervelt}},
  \bibinfo{author}{\bibfnamefont{P.~F.} \bibnamefont{Hopkins}},
  \bibnamefont{and} \bibinfo{author}{\bibfnamefont{A.~C.}
  \bibnamefont{Gossard}}, \bibinfo{journal}{Phys. Rev. Lett.}
  \textbf{\bibinfo{volume}{69}}, \bibinfo{pages}{506} (\bibinfo{year}{1992}).

\bibitem[{\citenamefont{Baranger et~al.}(1993)\citenamefont{Baranger, Jalabert,
  and Stone}}]{Baranger1993}
\bibinfo{author}{\bibfnamefont{H.~U.} \bibnamefont{Baranger}},
  \bibinfo{author}{\bibfnamefont{R.~A.} \bibnamefont{Jalabert}},
  \bibnamefont{and} \bibinfo{author}{\bibfnamefont{A.~D.} \bibnamefont{Stone}},
  \bibinfo{journal}{Phys. Rev. Lett.} \textbf{\bibinfo{volume}{70}},
  \bibinfo{pages}{3876} (\bibinfo{year}{1993}).

\bibitem[{\citenamefont{Chang et~al.}(1994)\citenamefont{Chang, Baranger,
  Pfeiffer, and West}}]{Chang1994}
\bibinfo{author}{\bibfnamefont{A.}~\bibnamefont{Chang}},
  \bibinfo{author}{\bibfnamefont{H.~U.} \bibnamefont{Baranger}},
  \bibinfo{author}{\bibfnamefont{L.}~\bibnamefont{Pfeiffer}}, \bibnamefont{and}
  \bibinfo{author}{\bibfnamefont{K.}~\bibnamefont{West}},
  \bibinfo{journal}{Phys. Rev. Lett.} \textbf{\bibinfo{volume}{73}},
  \bibinfo{pages}{2111} (\bibinfo{year}{1994}).

\bibitem[{\citenamefont{Sachrajda et~al.}(1998)\citenamefont{Sachrajda,
  Ketzmerick, Gould, Feng, Kelly, Delage, and Wasilewski}}]{Sachrajda1998}
\bibinfo{author}{\bibfnamefont{A.}~\bibnamefont{Sachrajda}},
  \bibinfo{author}{\bibfnamefont{R.}~\bibnamefont{Ketzmerick}},
  \bibinfo{author}{\bibfnamefont{C.}~\bibnamefont{Gould}},
  \bibinfo{author}{\bibfnamefont{Y.}~\bibnamefont{Feng}},
  \bibinfo{author}{\bibfnamefont{P.}~\bibnamefont{Kelly}},
  \bibinfo{author}{\bibfnamefont{A.}~\bibnamefont{Delage}}, \bibnamefont{and}
  \bibinfo{author}{\bibfnamefont{Z.}~\bibnamefont{Wasilewski}},
  \bibinfo{journal}{Phys. Rev. Lett.} \textbf{\bibinfo{volume}{80}},
  \bibinfo{pages}{1948} (\bibinfo{year}{1998}).

\bibitem[{\citenamefont{Marlow et~al.}(2006)\citenamefont{Marlow, Taylor,
  Martin, Scannell, Linke, Fairbanks, Hall, Shorubalko, Samuelson, Fromhold
  et~al.}}]{Marlow2006}
\bibinfo{author}{\bibfnamefont{C.}~\bibnamefont{Marlow}},
  \bibinfo{author}{\bibfnamefont{R.}~\bibnamefont{Taylor}},
  \bibinfo{author}{\bibfnamefont{T.}~\bibnamefont{Martin}},
  \bibinfo{author}{\bibfnamefont{B.}~\bibnamefont{Scannell}},
  \bibinfo{author}{\bibfnamefont{H.}~\bibnamefont{Linke}},
  \bibinfo{author}{\bibfnamefont{M.}~\bibnamefont{Fairbanks}},
  \bibinfo{author}{\bibfnamefont{G.}~\bibnamefont{Hall}},
  \bibinfo{author}{\bibfnamefont{I.}~\bibnamefont{Shorubalko}},
  \bibinfo{author}{\bibfnamefont{L.}~\bibnamefont{Samuelson}},
  \bibinfo{author}{\bibfnamefont{T.}~\bibnamefont{Fromhold}},
  \bibnamefont{et~al.}, \bibinfo{journal}{Phys. Rev. B}
  \textbf{\bibinfo{volume}{73}}, \bibinfo{pages}{195318}
  (\bibinfo{year}{2006}).

\bibitem[{\citenamefont{Timp et~al.}(1987)\citenamefont{Timp, Chang,
  Cunningham, Chang, Mankiewich, Behringer, and Howard}}]{Timp1987}
\bibinfo{author}{\bibfnamefont{G.}~\bibnamefont{Timp}},
  \bibinfo{author}{\bibfnamefont{A.~M.} \bibnamefont{Chang}},
  \bibinfo{author}{\bibfnamefont{J.~E.} \bibnamefont{Cunningham}},
  \bibinfo{author}{\bibfnamefont{T.~Y.} \bibnamefont{Chang}},
  \bibinfo{author}{\bibfnamefont{P.}~\bibnamefont{Mankiewich}},
  \bibinfo{author}{\bibfnamefont{R.}~\bibnamefont{Behringer}},
  \bibnamefont{and} \bibinfo{author}{\bibfnamefont{R.~E.}
  \bibnamefont{Howard}}, \bibinfo{journal}{Phys. Rev. Lett.}
  \textbf{\bibinfo{volume}{58}}, \bibinfo{pages}{2814} (\bibinfo{year}{1987}).

\bibitem[{\citenamefont{Olendski and Mikhailovska}(2005)}]{Olendski2005}
\bibinfo{author}{\bibfnamefont{O.}~\bibnamefont{Olendski}} \bibnamefont{and}
  \bibinfo{author}{\bibfnamefont{L.}~\bibnamefont{Mikhailovska}},
  \bibinfo{journal}{Phys. Rev. B} \textbf{\bibinfo{volume}{72}},
  \bibinfo{pages}{235314} (\bibinfo{year}{2005}).

\bibitem[{\citenamefont{Bird et~al.}(2003)\citenamefont{Bird, Akis, Ferry, {de
  Moura}, Lai, and Indelkofer}}]{Bird2003}
\bibinfo{author}{\bibfnamefont{J.~P.} \bibnamefont{Bird}},
  \bibinfo{author}{\bibfnamefont{R.}~\bibnamefont{Akis}},
  \bibinfo{author}{\bibfnamefont{D.~K.} \bibnamefont{Ferry}},
  \bibinfo{author}{\bibfnamefont{A.~P.~S.} \bibnamefont{{de Moura}}},
  \bibinfo{author}{\bibfnamefont{Y.-C.} \bibnamefont{Lai}}, \bibnamefont{and}
  \bibinfo{author}{\bibfnamefont{K.~M.} \bibnamefont{Indelkofer}},
  \bibinfo{journal}{Rep. Prog. Phys.} \textbf{\bibinfo{volume}{66}},
  \bibinfo{pages}{583} (\bibinfo{year}{2003}).

\bibitem[{\citenamefont{Elhassan et~al.}(2004)\citenamefont{Elhassan, Akis,
  Bird, Ferry, Ida, and Ishibashi}}]{Elhassan2004}
\bibinfo{author}{\bibfnamefont{M.}~\bibnamefont{Elhassan}},
  \bibinfo{author}{\bibfnamefont{R.}~\bibnamefont{Akis}},
  \bibinfo{author}{\bibfnamefont{J.~P.} \bibnamefont{Bird}},
  \bibinfo{author}{\bibfnamefont{D.~K.} \bibnamefont{Ferry}},
  \bibinfo{author}{\bibfnamefont{T.}~\bibnamefont{Ida}}, \bibnamefont{and}
  \bibinfo{author}{\bibfnamefont{K.}~\bibnamefont{Ishibashi}},
  \bibinfo{journal}{Phys. Rev. B} \textbf{\bibinfo{volume}{70}},
  \bibinfo{pages}{205341} (\bibinfo{year}{2004}).

\bibitem[{\citenamefont{Sadreev}(2004)}]{Sadreev2004}
\bibinfo{author}{\bibfnamefont{A.~F.} \bibnamefont{Sadreev}},
  \bibinfo{journal}{Phys. Rev. E} \textbf{\bibinfo{volume}{70}},
  \bibinfo{pages}{016208} (\bibinfo{year}{2004}).

\bibitem[{\citenamefont{von Klitzing et~al.}(1980)\citenamefont{von Klitzing,
  Dorda, and Pepper}}]{Klitzing1980}
\bibinfo{author}{\bibfnamefont{K.}~\bibnamefont{von Klitzing}},
  \bibinfo{author}{\bibfnamefont{G.}~\bibnamefont{Dorda}}, \bibnamefont{and}
  \bibinfo{author}{\bibfnamefont{M.}~\bibnamefont{Pepper}},
  \bibinfo{journal}{Phys. Rev. Lett.} \textbf{\bibinfo{volume}{45}},
  \bibinfo{pages}{494} (\bibinfo{year}{1980}).

\bibitem[{\citenamefont{Laughlin}(1981)}]{Laughlin1981}
\bibinfo{author}{\bibfnamefont{R.~B.} \bibnamefont{Laughlin}},
  \bibinfo{journal}{Phys. Rev. B} \textbf{\bibinfo{volume}{23}},
  \bibinfo{pages}{5632} (\bibinfo{year}{1981}).

\bibitem[{\citenamefont{Jalabert et~al.}(1992)\citenamefont{Jalabert, Baranger,
  and Stone}}]{Jalabert1992}
\bibinfo{author}{\bibfnamefont{R.~A.} \bibnamefont{Jalabert}},
  \bibinfo{author}{\bibfnamefont{H.~U.} \bibnamefont{Baranger}},
  \bibnamefont{and} \bibinfo{author}{\bibfnamefont{A.~D.} \bibnamefont{Stone}},
  \bibinfo{journal}{Phys. Rev. Lett.} \textbf{\bibinfo{volume}{68}},
  \bibinfo{pages}{3468} (\bibinfo{year}{1992}).

\bibitem[{\citenamefont{Wirtz et~al.}(2003)\citenamefont{Wirtz, Stampfer,
  Rotter, and Burgd{\"o}rfer}}]{Wirtz2003}
\bibinfo{author}{\bibfnamefont{L.}~\bibnamefont{Wirtz}},
  \bibinfo{author}{\bibfnamefont{C.}~\bibnamefont{Stampfer}},
  \bibinfo{author}{\bibfnamefont{S.}~\bibnamefont{Rotter}}, \bibnamefont{and}
  \bibinfo{author}{\bibfnamefont{J.}~\bibnamefont{Burgd{\"o}rfer}},
  \bibinfo{journal}{Phys. Rev. E} \textbf{\bibinfo{volume}{67}},
  \bibinfo{pages}{016206} (\bibinfo{year}{2003}).

\bibitem[{\citenamefont{Jacquod and Whitney}(2006)}]{Jacquod2006}
\bibinfo{author}{\bibfnamefont{P.}~\bibnamefont{Jacquod}} \bibnamefont{and}
  \bibinfo{author}{\bibfnamefont{R.~S.} \bibnamefont{Whitney}},
  \bibinfo{journal}{Phys. Rev. B} \textbf{\bibinfo{volume}{73}},
  \bibinfo{pages}{195115} (\bibinfo{year}{2006}).

\bibitem[{\citenamefont{Steinbach et~al.}(1996)\citenamefont{Steinbach,
  Martinis, and Devoret}}]{Steinbach1996}
\bibinfo{author}{\bibfnamefont{A.~H.} \bibnamefont{Steinbach}},
  \bibinfo{author}{\bibfnamefont{J.~M.} \bibnamefont{Martinis}},
  \bibnamefont{and} \bibinfo{author}{\bibfnamefont{M.~H.}
  \bibnamefont{Devoret}}, \bibinfo{journal}{Phys. Rev. Lett.}
  \textbf{\bibinfo{volume}{76}}, \bibinfo{pages}{3806} (\bibinfo{year}{1996}).

\bibitem[{\citenamefont{Henny et~al.}(1999)\citenamefont{Henny, Oberholzer,
  Strunk, and Sch{\"o}neberger}}]{Henny1999}
\bibinfo{author}{\bibfnamefont{M.}~\bibnamefont{Henny}},
  \bibinfo{author}{\bibfnamefont{S.}~\bibnamefont{Oberholzer}},
  \bibinfo{author}{\bibfnamefont{C.}~\bibnamefont{Strunk}}, \bibnamefont{and}
  \bibinfo{author}{\bibfnamefont{C.}~\bibnamefont{Sch{\"o}neberger}},
  \bibinfo{journal}{Phys. Rev. B} \textbf{\bibinfo{volume}{59}},
  \bibinfo{pages}{2871} (\bibinfo{year}{1999}).

\bibitem[{\citenamefont{Oberholzer et~al.}(2001)\citenamefont{Oberholzer,
  Sukhorukov, Strunk, Sch{\"o}nenberger, Heinzel, and
  Holland}}]{Oberholzer2001}
\bibinfo{author}{\bibfnamefont{S.}~\bibnamefont{Oberholzer}},
  \bibinfo{author}{\bibfnamefont{E.~V.} \bibnamefont{Sukhorukov}},
  \bibinfo{author}{\bibfnamefont{C.}~\bibnamefont{Strunk}},
  \bibinfo{author}{\bibfnamefont{C.}~\bibnamefont{Sch{\"o}nenberger}},
  \bibinfo{author}{\bibfnamefont{T.}~\bibnamefont{Heinzel}}, \bibnamefont{and}
  \bibinfo{author}{\bibfnamefont{M.}~\bibnamefont{Holland}},
  \bibinfo{journal}{Phys. Rev. Lett.} \textbf{\bibinfo{volume}{86}},
  \bibinfo{pages}{2114} (\bibinfo{year}{2001}).

\bibitem[{\citenamefont{Berry}(1981)}]{Berry1981}
\bibinfo{author}{\bibfnamefont{M.~V.} \bibnamefont{Berry}},
  \bibinfo{journal}{Europ. J. Phys.} \textbf{\bibinfo{volume}{2}},
  \bibinfo{pages}{91} (\bibinfo{year}{1981}).

\bibitem[{\citenamefont{Ferry and Goodnick}(1997)}]{Ferry1997}
\bibinfo{author}{\bibfnamefont{D.~K.} \bibnamefont{Ferry}} \bibnamefont{and}
  \bibinfo{author}{\bibfnamefont{S.~M.} \bibnamefont{Goodnick}},
  \emph{\bibinfo{title}{Transport in Nanostructures}}
  (\bibinfo{publisher}{Cambridge University Press}, \bibinfo{year}{1997}).

\bibitem[{\citenamefont{Drouvelis et~al.}(2006)\citenamefont{Drouvelis,
  Schmelcher, and Bastian}}]{Drouvelis2006}
\bibinfo{author}{\bibfnamefont{P.~S.} \bibnamefont{Drouvelis}},
  \bibinfo{author}{\bibfnamefont{P.}~\bibnamefont{Schmelcher}},
  \bibnamefont{and} \bibinfo{author}{\bibfnamefont{P.}~\bibnamefont{Bastian}},
  \bibinfo{journal}{J. Comp. Phys.} \textbf{\bibinfo{volume}{215}},
  \bibinfo{pages}{741} (\bibinfo{year}{2006}).

\bibitem[{\citenamefont{Rotter et~al.}(2003)\citenamefont{Rotter, Weingartner,
  Rohringer, and Burgd{\"o}rfer}}]{Rotter2003}
\bibinfo{author}{\bibfnamefont{S.}~\bibnamefont{Rotter}},
  \bibinfo{author}{\bibfnamefont{B.}~\bibnamefont{Weingartner}},
  \bibinfo{author}{\bibfnamefont{N.}~\bibnamefont{Rohringer}},
  \bibnamefont{and}
  \bibinfo{author}{\bibfnamefont{J.}~\bibnamefont{Burgd{\"o}rfer}},
  \bibinfo{journal}{Phys. Rev. B} \textbf{\bibinfo{volume}{68}},
  \bibinfo{pages}{165302} (\bibinfo{year}{2003}).

\bibitem[{Fus()}]{Fussnote1}
\bibinfo{note}{This notation is a priori not unique opposite to the rectangular
  billiard: the nodal lines in the oval are bent and if the wavefunction has a
  large amplitude close to the lead stubs, as e.g. in Fig.~\ref{abb_3}.c) and
  \ref{abb_3}.d), it relaxes towards the lead openings forming a locally more
  complicated nodal pattern. It is also not applicable to the circle
  ($\delta=0$) where the radial quantum number and the angular momentum are
  good quantum numbers. However, it describes very well the characteristics of
  the eigenstates for $\delta \ge 0.5$.}

\bibitem[{\citenamefont{Fano}(1961)}]{Fano1961}
\bibinfo{author}{\bibfnamefont{U.}~\bibnamefont{Fano}}, \bibinfo{journal}{Phys.
  Rev.} \textbf{\bibinfo{volume}{124}}, \bibinfo{pages}{1866}
  (\bibinfo{year}{1961}).

\end{thebibliography}
\end{document}